\documentclass[reprint,nofootinbib,amsmath,amssymb,amsfonts,aps,]{revtex4-1}

\usepackage{natbib}
\usepackage{color}
\usepackage{xcolor}
\usepackage{amsthm}   
\usepackage[normalem]{ulem}

\theoremstyle{plain}

  \theoremstyle{definition}

\usepackage[titletoc,toc,title]{appendix} 
\usepackage{graphicx}                     
\usepackage{dcolumn}                      
\usepackage{bm}                           
\usepackage[utf8]{inputenc}
\usepackage[
margin=2.5cm,
]{geometry}
\usepackage{amsmath}
\usepackage{amsfonts}
\usepackage{mathtools}
\usepackage{dsfont}
\usepackage{float} 
\usepackage{wrapfig} 
\usepackage[caption=false]{subfig}

\usepackage{bbold}
\usepackage{dsfont}
\usepackage{nccmath}
 
\usepackage{xcolor, soul}
\sethlcolor{red}

\usepackage{hyperref}
\hypersetup{
    colorlinks=true,
    linkcolor=black,
    filecolor=magenta,      
    urlcolor=blue}
\usepackage{cleveref} 

\usepackage[export]{adjustbox}

\usepackage[bottom]{footmisc}

\usepackage[colorinlistoftodos,prependcaption,textsize=tiny]{todonotes}

\usepackage{titlesec}

\titleformat{\paragraph}
{\normalfont\normalsize\bfseries}{\theparagraph}{1em}{}
\titlespacing*{\paragraph}
{0pt}{3.25ex plus 1ex minus .2ex}{1.5ex plus .2ex}

\begin{document}

\title{Transitioning from a bounce to $R^2$ inflation}

\author{Richard Daniel$^1$, Mariam Campbell$^2$, Carsten van de Bruck$^1$ and Peter Dunsby$^{2,3}$}
\affiliation{$^1$School of Mathematics and Statistics, University of Sheffield, Hounsfield Road, Sheffield S3 7RH, United Kingdom}%
\affiliation{$^2$Department of Mathematics and Applied Mathematics, Cosmology and Gravity Group,  University of Cape Town, Rondebosch, 7701, Cape Town, South Africa}
\affiliation{$^3$South African Astronomical Observatory, Observatory 7925, Cape Town, South Africa}

\date{May 19, 2023}
\begin{abstract}
Non-singular bouncing cosmologies are well--motivated models for the early universe. Recent observational data are consistent with positive spatial curvature and allow  for a natural collapsing and bouncing phase in the very early universe. Additionally, bouncing cosmologies have the potential to rectify conceptual shortcomings identified in the theory of inflation, such as the singularity problem. In this paper we present a classical bouncing model in the context of modified gravity, including an $R^2$-term in the action. We show that after the bounce, the universe enters naturally a period of inflation, driven by the $R^2$--term. We analyse the stability of the model and find that the scalaron assists the stability of the model.

\end{abstract}
\maketitle

\onecolumngrid

\newpage
\section{Introduction}
Inflation is a well motivated paradigm which makes a number of non--trivial predictions which have been found to be in agreement with a variety of cosmological observations. In particular it is in agreement with measurements of anisotropies in the cosmological microwave background (CMB) radiation \cite{Planck:2018vyg} and the fact that the universe is spatially nearly flat. Importantly, it provides a self-consistent causal explanation for the origin of large scale structures (galaxies \& clusters of galaxies) via quantum fluctuations that are stretched to cosmological scales, providing the seeds for structure formation \cite{Baumann:2009ds}.  Inflation was proposed as a mechanism to relieve initial condition problems that arose with the original Hot Big Bang model. Despite curing these initial conditions problems, the theory of inflation still depicts a universe originating from a singular point \cite{1970RSPSA.314..529H, Borde:2001nh} \footnote{Assuming the validity of General Relativity (GR).}. The idea of replacing the singular origin, the Big Bang, with a non-singular transition, a bounce, is not a new one and we refer to extensive reviews on this topics, such as e.g. \cite{Battefeld:2014uga, Novello:2008ra, Brandenberger:2016vhg}. A model with a single bounce does not necessarily cure the initial condition (or singularity) problem, but simply pushes the solution back to an earlier state. Many different types of cosmological bouncing models have been proposed, e.g. using extra dimensions and branes \cite{Khoury:2001bz, Khoury:2001wf,Mukherji:2002ft, Shtanov:2002mb, Lehners:2008vx, Nojiri:2022xdo}, modifications to gravity or the kinetic terms \cite{Easson:2011zy,Qiu:2015nha, Ijjas:2016tpn, Nojiri:2017ncd,Agrawal:2022lhh,Dobre:2017pnt,Nojiri:2019lqw, Odintsov:2021yva}, or simply arising as a fundamental starting point in quantum cosmologies \cite{Pinto-Neto:2021gcl}. One of the early attempts at avoiding the initial singularity was presented in \cite{PhysRevD.7.2357}, by considering a closed Robertson-Walker geometry. They found that a violation of the energy conditions due to the quantum effects in their model results in a transition from a Friedmann-like collapse to a Friedmann-like expansion. However, in their solution the universe will eventually collapse after completing a cycle, and their model cannot show if quantum effects would always avoid a gravitational collapse. Many of these models propose an alternative to inflation and aim to predict the observed anisotropies of the CMB within these alternative scenarios \cite{Gungor:2020fce, Brandenberger:2008zz, Lyth:2001nq}. Unfortunately, many alternatives to inflation will suffer from a fine-tuning problem, due to instabilities in the energy components driving the bounce. Nevertheless, these models provide a starting point to formulate a more complete theory of the universe, e.g. in which the universe undergoes cycles. 

Another potential conceptual weakness of inflation is the somewhat arbitrary initial conditions to get inflation started \cite{Brandenberger:2016uzh, Linde:2017pwt}. The idea of an emergent universe proposed in \cite{2004CQGra..21..223E} may provide a scenario of a singularity-free inflationary universe. They do not replace the initial singularity with a bounce, but rather consider a universe that emerges from an Einstein static state far in the past, such that quantum gravity effects are negligible before entering the inflationary phase. This idea could, in principle, be used to resolve any issues with an inflationary induced bounce by pushing the  initial singularity to an earlier cosmological time. This allows for an emergent universe originating from an Einstein static state, transitioning through a bounce and into an inflationary epoch. 

Nevertheless, theories with a non--singular bounce usually require either a violation of the null--energy condition or non--zero curvature \cite{Falciano:2008gt, Shabani:2017kis}. A number of papers have studied bouncing cosmologies in the context of scalar--tensor theories, see e.g. \cite{Boisseau:2016pfh,Polarski:2021azv,Mukherjee:2022yvx}. The model we will discuss in this paper relies on the assumption that our universe is spatially closed and takes its inspiration from \cite{Gungor:2020fce}\footnote{Recent analyses of Planck data are consistent with a small positive curvature \cite{Planck:2018vyg,DiValentino:2019qzk,Yang:2022kho,Handley:2019tkm}. In our model the spatial curvature at the present can indeed be very close to zero.}. It relies on a modified gravity theory with a non--minimally coupled scalar field. The scalar field is assumed to sit in the minimum of an effective potential initially. Inflation after the bounce is driven by corrections to the Einstein--Hilbert action, which we assume to be of the form of $R + R^2$ gravity\footnote{Bouncing cosmologies in these type of modified gravity theories have been discussed in \cite{PhysRevD.36.1607, 2006CQGra..23.1913C}.}.
We find that for a large range of parameter, the scalar field rolls down towards the minimum of an effective potential, thereby triggering a bounce, while the scalaron is driven up its potential energy, resulting in a period of Starobinsky--inflation. Our model is very similar to \cite{Gungor:2020fce}, with the addition of the dynamics of the scalaron field. The main result of our paper is that inflation driven by the $R^2$--terms happens naturally in such a setup and we provide  evidence that the model is stable for a large region of parameter space. 

The paper is organised as follows: In the next Section we present the model and derive the relevant equations of motion. In Section III, we present the initial conditions for the fields, discuss the dynamics of the bounce and the period of inflation afterwards. In Section IV we discuss the evolution of cosmological perturbations. The equations for the field and metric perturbations are studied numerically. Our conclusions can be found in Section V. 
\section{The Model}
We assume a closed, isotropic and homogeneous background described by the Friedmann--Robertson--Lemaitre--Walker (FRLW) metric  
\begin{equation}
    ds^2 = -dt^2 + a(t)^2\left[ \frac{dr^2}{1- Kr^2} + r^2\left(d\theta^2 + \sin^2\theta d\phi^2\right) \right]~,
\end{equation}
with $K>0$. Our theory is an extension of Starobinsky's $R^2$ model \cite{Starobinsky:1980te} with an additional scalar field non-minimally coupled to gravity, 
\begin{equation}
    {\cal S} = \int d^4x\sqrt{-g}
    \left[ \frac{1}{2}\left(M_{\rm Pl}^2 - \alpha \phi^2 \right)R + \frac{1}
    {2}AR^2 - \frac{1}{2}(\nabla \phi)^2 - V(\phi)
    \right].
    \label{eqn full action}
\end{equation}
Here $\alpha$ and $A$ are constants parameterising the coupling of the scalar field to GR and the modification of Starobinsky's term respectively. We follow \cite{Gungor:2020fce} and choose the potential to be of the form 
\begin{equation}
    V =  \frac{m^2}{2}\phi^2 + \frac{\beta}{3}\phi^3 + \frac{\lambda}{4}\phi^4.
    \label{eqn potential}
\end{equation}
We map the action \eqref{eqn full action} to a bi--scalar--tensor theory by defining $f(\phi,R) = \frac{1}{2}\left(M_{\rm Pl}^2 - \alpha \phi^2 \right)R + \frac{1} {2}AR^2$. Our additional scalar degree of freedom is then defined in the standard way setting $\psi = \partial f/\partial R \equiv f_R$. This allows us to write the Ricci scalar and the function $f(R,\phi)$ in terms of the two fields as  
\begin{align}
\begin{aligned}
    R = \frac{1}{A}\left(\psi - \frac{M_{\rm Pl}^2 - \alpha \phi^2}{2}\right),~~~~
    f = \psi R  - \frac{\left[\psi - \frac{1}{2}(M_{\rm Pl}^2 - \alpha \phi^2)\right]^2}{2A}.
\end{aligned}
\label{eqn R f}
\end{align}

The field equations are obtained from the action by taking the variation with respect to the metric. We obtain 
\begin{equation}
    \psi R_{\mu\nu} - \frac{f g_{\mu\nu}}{2} - (\nabla_\mu\nabla_\nu - g_{\mu\nu}\Box)\psi = T_{\mu\nu}^{(\phi)}~.
    \label{eqn eom metric}
\end{equation}
Taking the trace of the last equation and making use of equations (\ref{eqn R f}) we determine the evolution equation for $\psi$, \begin{equation}
    \Box \psi = \frac{1}{3}\left[\frac{M_{\rm Pl}^2-\alpha\phi^2}{2A}\left(\psi - \frac{1}{2}(M_{\rm Pl}^2-\alpha\phi^2)\right) +  T \right], 
\end{equation}
where $T$ is the trace of the energy momentum tenor of the $\phi$ field. In a closed FRLW spacetime, the equations of motions for the fields $\psi$ and $\phi$ read 
\begin{align}
    \ddot{\psi} + 3H\dot{\psi} &= \frac{1}{3}\left[\frac{M_{\rm Pl}^2-\alpha\phi^2}{2A}\left( \frac{1}{2}(M_{\rm Pl}^2-\alpha\phi^2)- \psi \right) +  ( \rho -3P) \right]
    \label{eqn eom psi}
    \\
    \ddot{\phi} + 3H\dot{\phi} &= - V_\phi - \alpha\phi R.
    \label{eqn eom phi jordan}
\end{align}
The Friedmann equations are given by  
\begin{align}
    H^2 + \frac{K}{a^2} &= \frac{\psi R - f}{6\psi} + \frac{\rho}{3\psi} - H\frac{\dot{\psi}}{\psi},
    \label{eqn H2 jordan}
    \\
    \dot{H} - \frac{K}{a^2} &= \frac{H\dot{\psi}}{2\psi}  -\frac{\ddot{\psi}}{2\psi} - \frac{ (\rho + P)}{2\psi},
    \label{eqn another Hdot jordan}
    \\
    \dot{H} &= \frac{R}{6} - 2H^2 - \frac{K}{a^2}.
    \label{eqn Hdot jordan} 
\end{align}
In what follows, we will study a bouncing cosmological scenario, which is an extension of the model presented in \cite{Gungor:2020fce}. We aim to use the bounce to set the initial conditions for a subsequent inflationary epoch, driven by the $R^2$ term. 
\section{Bounce dynamics}
In what follows, we will be working in the Jordan frame to analyse the cosmological dynamics. An analysis in the Einstein frame is also possible, but we will refrain here from analysing the dynamics in that frame. Given the potential \eqref{eqn potential}, we utilise the picture set forth by \cite{Gungor:2020fce}: pre--bounce we assume a slowly contracting universe dominated by dark energy. During this time, $\phi$ has settled in the false vacuum, thereby providing initial conditions for our fields. The location of the false vacuum is given by {the} effective potential for $\phi$. From \eqref{eqn eom phi jordan} we find extrema located at $\phi=0$ (the true vacuum) and at  
\begin{equation}
    \phi_\pm = \frac{-\beta \pm \sqrt{\beta^2 - 4\lambda (m^2 + \alpha R)}}{2\lambda}.
    \label{eqn phi inital}
\end{equation}

\begin{figure}
    \centering
    \includegraphics[width = \textwidth]{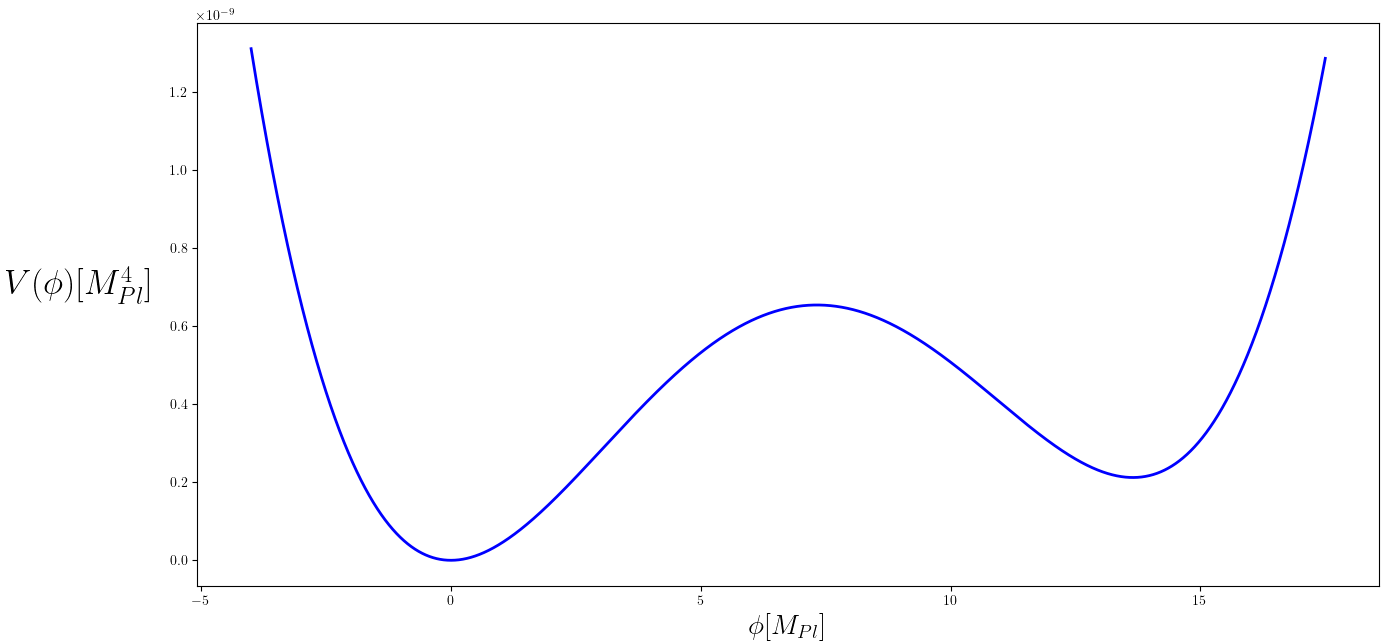}
    \caption{The potential \cref{eqn potential} for $\phi$, as studied in \cite{Gungor:2020fce}. The initial conditions of our scalar field will be determined by the location of the local minimum given by the choice of parameters of the potential. The parameters for this potential are $m=10^{-6}M_{\rm Pl},~\beta=-\sqrt{4.4\lambda}m$ and $\lambda = 10^{-12}$.}
    \label{fig potential}
\end{figure}

The negative solution corresponds to the local maximum shown in \cref{fig potential} and the plus solution is the semi-stable local minimum. To ensure that we have a local minimum, we require $\beta^2>4\lambda m^2$. Furthermore, we assume that the potential is positive at $\phi_-$ which constrains $\beta^2<4.5\lambda m^2$. It is worth noting that a deeper local minimum corresponds to a larger $\beta^2$.

Before the bounce, the field has settled at the initial value $\phi_i = \phi_+$. For $t\rightarrow-\infty$, $R$ will be very small compared to the mass of the field by considering a slowly contracting universe, allowing us to disregard $R$ in \eqref{eqn phi inital}. 

 In order to guarantee that the value of $R$ is initially negligible, we set the parameter $A$ to be large \footnote{It is worth noting $A$ has units of $[mass]^{-1/2}$, and in standard Starobinsky inflation, $M_\text{Pl}^2\gg A^{-1}$.} compared to the fields, as seen in eqn. \eqref{eqn R f}.

Since $R$ is initially negligible by forcing $A$ we place $\phi$ at the local minimum, we will also assume the scaleron is initially at rest. We obtain from eqn. \eqref{eqn eom psi}
\begin{equation}
    \psi_{\rm min}(\phi) = \frac{8AV(\phi_i)}{M_{\rm Pl}^2-\alpha\phi_i^2} + \frac{1}{2}(M_{\rm Pl}^2-\alpha\phi_i^2),
    \label{eqn psi min}
\end{equation}
where we will set the initial value of the scaleron such that $\psi_i = \psi_{\rm min}(\phi_i)$.

If $\phi$ evolves very slowly, we can immediately see that $\psi$ will track $\phi$. It is also clear that we cannot allow $\alpha\phi^2=M_{\rm Pl}^2$, otherwise the scaleron will be unbounded. This identifies a constraint on our model. Without loss of generality we assume a natural range for $\phi$ up to $O(10~M_{\rm Pl})$. Therefore we set $\alpha=10^{-3}$ throughout this paper, unless stated otherwise. 

Using the initial conditions provided above, we find 
\begin{eqnarray}
    V(\phi_i) &\approx& O(m^4)
    \\
    R &\approx& \frac{8V}{M_{\rm Pl}^2 - \alpha\phi^2} \approx O\left( \frac{m^4}{M_{\rm Pl}^2}\right)
    \\
    H^2 &\approx& \frac{V}{3(M_{\rm Pl}^2-\alpha\phi^2)}\approx O\left(\frac{m^4}{M_{\rm Pl}^2}\right)
    \\
    \dot{H} &\approx& 0
\end{eqnarray}
\subsubsection*{Approaching the bounce}
In \cite{Gungor:2020fce} it was argued that the field $\phi$ does not need to be fine tuned to allow a stable solution in the past for this given potential. This was shown by performing an adiabatic shift to first order around $V(\phi_i))$:
\begin{equation}
    \phi(t) \approx \phi_i + \frac{\dot{\phi}_i}{\omega}e^{-\frac{3H(t-t_i)}{2}}\sin\left(\frac{\omega(t - t_i)}{2}\right),
\end{equation}
where
\begin{equation}
    \omega = \sqrt{\phi_i\frac{(\beta^2 - 4\lambda m^2)^{1/2}}{\lambda} - H^2}.
\end{equation}
Initially, $H$ is constant (and negative), such that $\dot{H}\approx 0$. Using this assumption as well as equations \eqref{eqn Hdot jordan} and \eqref{eqn phi inital}, we can approximate 
\begin{equation}
    \dot{\phi_i} = \frac{\dot{R}}{\beta\sqrt{1 - \frac{4\lambda m^2}{\beta^2}}} = - 12\frac{\alpha H K}{a^2\beta\sqrt{1 - \frac{4\lambda m^2}{\beta^2}}}
\end{equation}
Combining this with the adiabatic shift, $\phi$ initially behaves as, 
\begin{equation}
    \phi \approx \phi_i + 12 \frac{\alpha|H|K}{\bar{\omega}}e^{\frac{-|H|(t-t_i)}{2}}\sin\left(\frac{\omega(t - t_i)}{2}\right),
\end{equation}
where 
\begin{equation}
    \bar{\omega} = \sqrt{\phi_i\frac{(\beta^2 - 4\lambda m^2)}{\lambda} - H^2(\beta^2 - 4\lambda m^2)^{1/2}},
\end{equation}
and we have used the fact that ${H}$ is a constant. 

It is clear that $\phi\rightarrow\phi_i$ as $t\rightarrow-\infty$. On the other hand we see that a small perturbation away from the minimum will eventually become important due to the anti-Hubble damping term,
and is dependent on both $H$ and $\bar{\omega}$, both of which are given by the parameters of the model. This hints that $\phi$ determines the dynamics of the bounce irrespective of $\psi$, assuming $R$ is set to be negligible at times well before the bounce. We can show this more explicitly by considering the dynamics approaching the bounce. The beginning and ending of the bouncing epoch is when the universe switches from an accelerated to a decelerated contraction phase, or vise versa, characterised by $\dot{H}=0$, leading to
\begin{equation}
    \frac{K}{a^2} = - \frac{H\dot{\psi}}{2\psi} + \frac{\ddot{\psi}}{2\psi} + \frac{ (\rho + P)}{2\psi}.
\end{equation}
This corresponds to $H$ reaching its extreme value $H_{\rm min/max}$,
\begin{equation}
    H_{\rm min/max}^2 = \frac{\psi R - f}{6\psi} - \frac{ (\rho + 3P)}{6\psi} + \frac{|H|\dot{\psi}}{2\psi} + \frac{\ddot{\psi}}{2\psi}. 
\end{equation}
At beginning of the bouncing epoch we have that $H\approx H_{\rm min}$. If we assume that the fields are well behaved and they do not rapidly diverge in the collapsing epoch to avoid the singularity ($\ddot{\psi},~ \ddot{\phi}< H\dot{\psi},~ H\dot{\phi}$), we can assume the scaleron will simply trace $\phi$ according to \eqref{eqn psi min}. This allows us to determine the Hubble parameter at the beginning of the bouncing epoch, 
\begin{equation}
    H_{\rm min}^2 \approx \frac{2V}{3 M_{\rm Pl}^2} + \frac{ V - \dot{\phi}^2}{3\psi(\phi)} + \frac{\dot{\phi}^2}{54\psi(\phi)}.\label{eqn H_min}
\end{equation}
Here we have made use of the assumed the hierarchy $\psi_i>M_{\rm Pl}^2\gg \alpha\phi_i^2$ to allow for inflation, explained in \cref{sec Inflation}. It is clear from \eqref{eqn H_min} that the dynamics before the bounce is determined by $\phi$. 

The scalar field $\phi$ will be displaced from the local minimum at some point during the collapsing phase, e.g. because of the presence of (small) perturbations. Perturbations will force the field value to slowly grow; however this generally happens very slowly and leads to two scenarios: either the fields remain being trapped in their local minima, expressed in \cref{fig backgound dynamics jordan}, or the fields escape their minima, a scenario that can be seen in \cref{fig background bounce}. The coupling of $\phi$ to the Ricci scalar introduces a time--varying effective potential, which can be controlled by $\alpha$, and lead to the local minimum to vanish. The $\phi$--field will roll towards the global minimum at $\phi=0$. This is the scenario discussed in \cite{Gungor:2020fce}. The inclusion of an additional degree of freedom here (the scalaron) allows $R$ to grow sufficiently, but not to the extent that the local minimum vanishes. Further below we discuss the different outcomes, the trapped $\phi$--field and the scenario where the potential is shallower, allowing the $\phi$--field to roll to 0. 

\subsection{Numerical analysis}
We perform a numerically analysis to determine the evolution of the fields and the evolution of the universe. To this end, we integrate the field equations \cref{eqn eom psi,eqn eom phi jordan,eqn H2 jordan,eqn Hdot jordan}, with initial conditions given by \cref{eqn phi inital,eqn psi min} and with both fields starting at rest. Due to the long time of integration, the time has been re-scaled by $m/M_{\rm Pl}$. For ease we also plot the e-fold, defined as $N = \log(a)$ normalised at the time of the bounce, to illustrate the transition. We have checked the validity of our numerical solutions by ensuring that the conditions for a bounce have been met (see \cref{sec bounce cond} for details) and verified that the Ricci scalar $$R = 6\left( \dot H + 2H^2 + \frac{K}{a^2}\right)$$ is in agreement with eq. \cref{eqn R f}. From our analysis we establish three possible outcomes on how the universe can evolve in this scenario. These outcomes are controlled by a choice of parameters that determine the evolution of $\phi$ and correspondingly the Ricci scalar at the bounce. 

\begin{itemize}
\item \underline{{$\phi\approx 0$} at or near the bounce}. The field $\phi$ is able to escape the false vacuum before the bounce and rolls towards the true minimum. This scenario can be obtained by forcing the evolution of $R$ such that $R \geq R_{crit}$, at which point the local minimum cease to exist and the field $\phi$ starts to evolve. This was explored by \cite{Gungor:2020fce} for a single field, where they were able to avoid the singularity by forcing the local minimum to be very close to the global minimum. Another way to allow the scalar field to freely oscillate is to construct the potential to have a very shallow and small barrier. This allows for the anti-friction term in \eqref{eqn eom phi jordan} to be the initial dominating term removing any oscillations. Unless severely fine tuned, the fields in this scenario will exhibit the standard divergent nature of bouncing mechanisms quickly leading to a singularity \cite{Gordon:2002jw, Falciano:2008gt}. Therefore, we do not explore this scenario further.

\item{\underline{$\phi\approx\phi_i$ at bounce:}} The field $\phi$ is never significantly displaced from $\phi_i$ the fields will remain trapped in their false vacuum state and $R$ never reaches $R_{crit}$ and does not evolve sufficiently to alter the potential. Therefore after the bounce $\phi$, $\psi$, and consequently $R$ will settle back in their initial conditions. This scenario is depicted in \cref{fig backgound dynamics jordan}, clearly showing a return to initial conditions after the bounce. This then leads to an eternal inflation scenario with the dark energy of the previous universe continuing to dominate. This can be due to either parameter choices creating a very steep and deep false vacuum, trapping the field through the bouncing epoch. Another reason is the chosen parameters do not allow $R$ to vary. $R$ can be set to have a minimal evolution through a parameter choice, such as setting the scalaron mass very high or reducing the coupling between $\phi$ and gravity. In both cases the field only undergoes minor oscillations as the spatial curvature dominates. This means that the effective potential remains unchanged ($V,_{\phi}\gg \alpha\phi R$). The velocity term never dominates the right hand side of \eqref{eqn eom psi}, hence $\psi \approx \psi_{min}$, tracing the $\phi$ field. This creates a symmetric bounce as illustrated in \cref{fig backgound dynamics jordan} as the scaleron is determined completely by the evolution of $\phi$. Since we end up in an eternally inflating universe, we focus our attention to scenarios with a bounce resulting in standard inflation.

\begin{figure}[t]
    \centering
    \includegraphics[width=\textwidth]{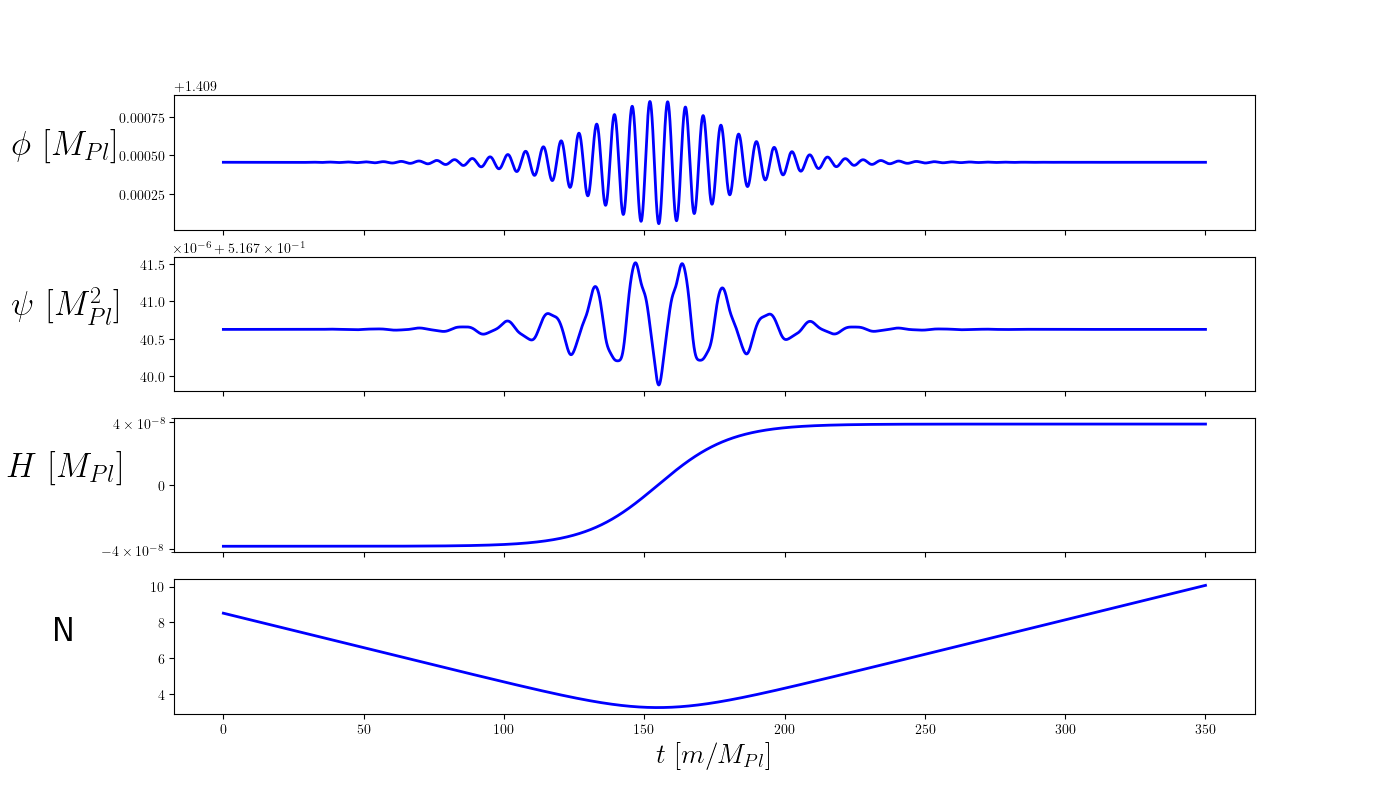}
    \caption{Behaviour of the fields whilst the fields are trapped in the false vacuum. We see that the field behave initially as expected in a collapsing universe, oscillating around their local minimum growing in amplitude. The bounce is then caused when the spatial curvature dominates, switching to a expanding universe and the fields become damped. The parameters used are $m=10^{-6}\text{M}_{\text{Pl}},~\beta = -\sqrt{4.49\lambda}m,~\lambda = 10^{-12},~ \alpha = 10^{-3}, ~V_0=0$, $A= 10^{12},~a = 10^2, K= m^2$ which in turn determine the initial conditions, $\phi_i$ and $\psi_{i}$}. 
    \label{fig backgound dynamics jordan}
\end{figure}

\item \underline{{$0<\phi\lesssim\phi_-$ at the bounce}:} In the final scenario, the Ricci scalar can evolve sufficiently such that $\phi$ is displaced but the false vacuum is not removed (i.e. the value of the Ricci scalar remains below $R_{\rm crit}$). In this case, $\phi$ exhibits growing oscillations due to the fact that the effective potential changes its form, eventually allowing the field to leave the local minimum before or as the bounce occurs, but remaining displaced from the global minimum at the time of the bounce, shown in \cref{fig background bounce}. The change of the potential is provides $\phi$ with enough kinetic energy to overcome the barrier ($|H|\dot{\phi}^2>V(\phi_-)$). In this scenario, the period of inflation following the bounce is initially driven by both fields. The field $\phi$ will always settle at the origin before $\psi$ resulting in a period of standard single field inflation driven by $\psi$, the behaviour of which can be seen in \cref{fig background psi_inflaton}. This epoch of inflation can then be constructed to last much longer than $60$ e-folds. Our choice of parameters forces the $\phi$-field to settle immediately, resulting in an almost entirely single field inflation. We present more details in \cref{sec Inflation}. 

\begin{figure}[t]
    \centering
        \includegraphics[width=\textwidth]{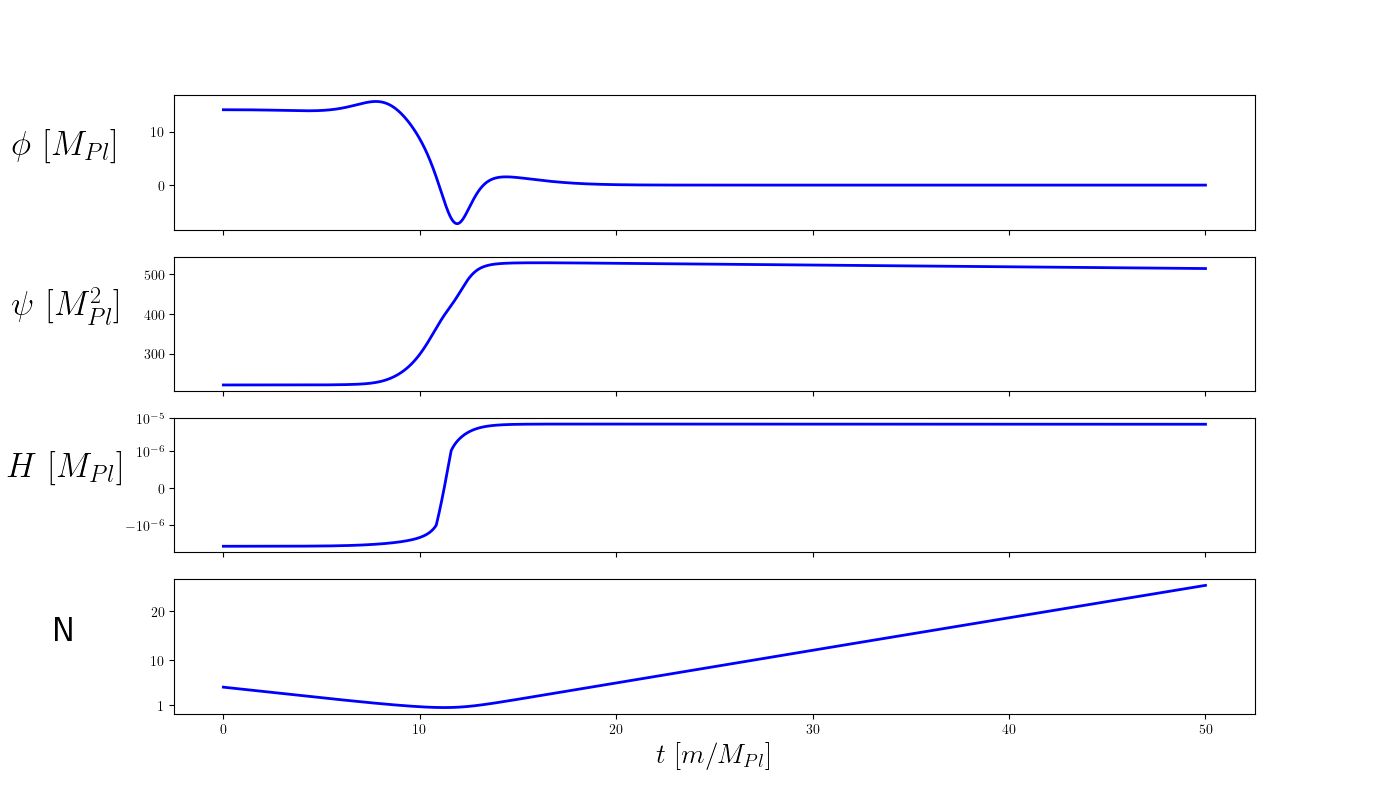}
    \caption{The dynamics of the fields provided they are able to leave the local minimum but the potential has not been sufficiently warped by the evolution of $R$. Although not realistic for our current universe, it illustrates the dynamics that creates slow-roll inflation desired. As the bounce occurs and $\phi$ escapes the local minimum, $\psi$ is driven up its potential. The parameters used are $m=10^{-5}\text{M}_{\text{Pl}},~\beta = -\sqrt{4.49\lambda}m,~\lambda = 10^{-12},~ \alpha = 10^{-3}, ~V_0=0$, $A= 10^{12}$, and $K = m^2$ which in turn determine the initial conditions, $\phi_i$ and $\psi_{i}$.  }
    \label{fig background psi_inflaton}
    \label{fig background bounce}
\end{figure}

\end{itemize}

We assume the fields to be in a slow rolling regime ($\ddot{\phi}, \ddot{\psi}\ll H\dot{\phi},H\dot{\psi}$) corresponding to the fields evolving slowly and $\phi$ gradually escaping the false vacuum. We can then approximate, using Eq.\eqref{eqn eom psi},
\begin{equation}
    3H\dot{\psi} \simeq \frac{M_{\rm Pl}^2(\frac{1}{2}M_{\rm Pl}^2 - \psi)}{6A} - \frac{\dot{\phi}^2 + 4V}{3} 
    \label{eqn psi approx}
\end{equation}
Given that we are motivated to find inflation after the bounce, we set the parameters such that $\psi_i>2M_{\rm Pl}^2$. Therefore in a collapsing universe, we see that $\psi$ is driven up its effective potential to larger values because its time--derivative is given by 
\begin{equation}
    \dot{\psi}_c \simeq \frac{\frac{M_{\rm Pl}^2 \psi}{A} + 2\dot{\phi}^2 + 8V}{18|H_c|}
    \label{eqn psi_c}>0,
\end{equation}
due to the fact the right hand side will remain positive. Therefore, while $\phi$ is driven towards zero, the potential energy of $\psi$ will become more relevant at time progresses. Hence, we arrive at a situation in which the bounce naturally  produces the initial conditions for inflation, driven by the $\psi$--field. This period of inflation begins when $H$ reaches its maximum value. We can use \eqref{eqn H_min} to determine the value of $\psi$ when inflation begins. For simplicity we assume that $H_{\rm min}$ is reached when $\phi = \phi_-$ and the potential is dominating \eqref{eqn H_min}. Therefore, we can equate $H_{\rm min}^2=H_{\rm max}^2$,  
\begin{equation}
    \frac{2V(\phi_-)}{3M_{\rm Pl}^2} = \frac{\psi R - f}{6\psi} - \frac{H\dot{\psi}}{\psi}
\end{equation}
using \eqref{eqn eom psi} and the definition of $f$ this results in 
\begin{equation}
    \psi_{\rm inf} = \frac{4AV(\phi_-)}{3M_{\rm Pl}^2} - \frac{M_{\rm Pl}^2}{18},
    \label{eqn psi inflation}
\end{equation}
where the subscript {\it inf} denotes the start of inflation. This calculation relies on the assumption of a symmetrical bounce. Due to the presence of the coupling of $\phi$ to the Ricci scalar as well as the scalaron, the bounce will not be symmetrical. However, given our choice of parameters, we have numerically verified that the scaleron plays a very minor role during the bounce. Therefore, the approximation $H_{\rm min}^2\approx H_{\rm max}^2$ is reasonable to find an approximation for $\psi_{\rm{inf}}$. Nevertheless, the scenario in which $H^2_{\rm max}>H^2_{\rm min}$ is more realistic. In this case the $\psi_{\rm inf}$ will be smaller, resulting in a shorter period of inflation. If $H^2_{\rm min}>H^2_{\rm max}$, $\psi_{\rm inf}$ is larger, resulting in a longer period of inflation.

\subsection{Resulting Inflation}\label{sec Inflation}
\begin{figure}[t]
    \centering
    \includegraphics[width = \textwidth]{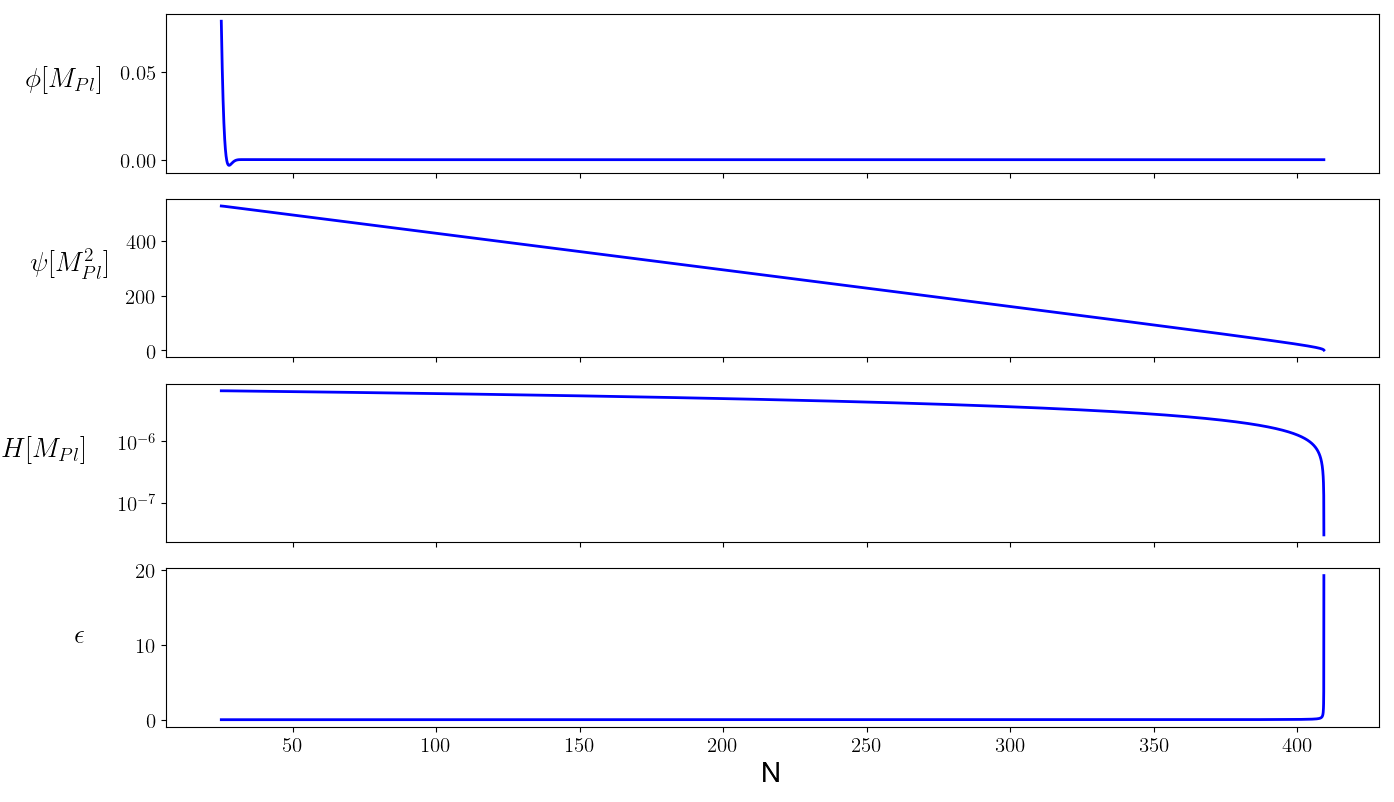}
    \caption{The evolution of the fields during inflation, using the same parameters as in \cref{fig background bounce}. We are using the e-fold number (measured after the bounce) as time variable in these figures.}
    \label{fig inflation}
\end{figure}

As stated before, in the scenario in which $\phi$ settles at 0 after the bounce, inflation is driven by the scalaron, with initial conditions provided at the end of the bouncing epoch. The slow-roll parameter becomes
\begin{equation}
    \epsilon_v = \frac{M_{\rm Pl}^2}{8 \psi^2}\left(\frac{\frac{M_{\rm Pl}^2}{2} - \psi}{M_{\rm Pl}^2 - \psi} \right)^2.
\end{equation}
Setting $\epsilon_v =1$ we can determine the end of inflation occurs when $\psi = 2M_{\rm Pl}^2$. Therefore we require to have $\psi_i>2M_{\rm Pl}^2$ at the beginning of that period.
Integrating our slow roll parameter allows us to determine the initial conditions of inflation for $\psi$ to achieve at least 60 e-folds of inflation. 
\begin{equation}
    N = \int^\psi_{2M_{\rm Pl}^2} \sqrt{\frac{2}{M_{\rm Pl}\epsilon_v}} 
\end{equation}
Using the same parameters as in Fig \cref{fig background psi_inflaton}, we find that $\psi_{inf}>16 M_{\rm Pl}^2$ to provide at least 60 e-folds of inflation driven by $\psi$. We can then use this value and Eq. \eqref{eqn psi inflation} to constrain the potential and the initial conditions of $\phi$. We leave the detailed analysis for future work. In  \cref{fig inflation} we show the evolution of the fields during inflation. 
\section{Cosmological Perturbations}\label{sec perturbations}
We now turn our attention to determining whether our model exhibits instabilities. We do this by considering the  evolution of cosmological perturbations at linear order, decomposing into scalar, vector and tesnor perturbations. In this Section we set $M_{\rm Pl} = 1$, unless stated otherwise.  

Focusing on the scalar branch, working in the longitudinal gauge, the metric at first order scalar perturbation is \cite{Mukhanov:1990me}\footnote{The scalar perturbations can also be derived using the variational approach as illustrated in \cite{PhysRevD.28.1809}.} 

\begin{equation}
    ds^2_{(s)} = 
    -(1+ 2{\Phi})dt^2 + a^2 (1 - 2{\Psi})\gamma_{ij} dx^i dx^j,
\end{equation}
where $\Phi$ and $\Psi$ are metric perturbations and $\gamma_{ij}$ is the metric on constant time hypersurfaces. The gauge invariant formalism for modified gravity is detailed in e.g. \cite{DeFelice:2010aj, Hwang:2001qk, Bardeen:1980kt} and references therein. The perturbed Einstein equation reads\footnote{Often in the literature in modified gravity the above equations will have a replacement of $\psi\rightarrow F$ \cite{Hwang:1996bc}.}  \cite{Hwang:1996bc}
\begin{align}
    3H \left(\dot{\Psi } +H\Phi\right)& +\frac{k^2-3 K }{a^2}\Psi =-\delta \rho,\label{eqn EFE pert energy}
    \\
    \dot{\Psi} + &H\Phi =  - \delta q, \label{eqn EFE pert momentum}
    \\
    3\left(\ddot{\Psi}+\dot{ H}\Phi -H\dot{\Phi} \right) +6 H & \left(\dot{\Psi }+H\Phi\right) + \Phi  \left(3 \dot{H}-\frac{k^2}{a^2}\right) = -\delta X, \label{eqn EFE pert differential}.
    \\
    \Psi - \Phi = \frac{\delta \psi}{\psi}, \label{eqn pert non-diag relation}
\end{align}
where we have defined 
\begin{align}
    \delta \rho &= \frac{1}{2\psi}\left[3 \dot{\psi} \left(\dot{\Psi }+H\Phi\right) - \left(\dot{\phi}^2 +3V\right)\Phi  -3 H \dot{\delta \psi}+\dot{\delta \phi } \dot{\phi} + \delta \psi \left(3 \dot{H} -\frac{k^2}{a^2}+3 H^2\right)
    +\delta \phi  \left(V_{\phi }-\frac{f_{\phi }}{2}\right)\right],
    \label{eqn delta rho}
    \\
    \delta q &= \frac{ \dot{\delta \psi}+\delta \phi  \dot{\phi }-H\delta \psi -\dot{\psi}\Phi}{2\psi},
    \label{eqn delta q}
    \\
    \delta X &= \frac{1}{2 \psi} \left[3 \dot{\psi} \dot{\Phi } + 3 \left(\dot{\Psi}+ H\Phi\right) \dot{\psi}+\Phi  \left( 6 \ddot{\psi}+3 H \dot{\psi}+4 \dot{\phi}^2 \right)
    \right.\nonumber 
    \\
     &~~~~~~~\left.+ \delta \psi \left(\frac{6 K-k^2}{a^2}+6 H\right) -\delta \phi  \left(f_{\phi }-2 V_{\phi }\right)
    -3 H \dot{\delta \psi}-3 \ddot{\delta \psi}-4 \dot{\delta \phi } \dot{\phi }\right].
    \label{eqn delta x}
\end{align}
The perturbed Klein-Gordon equations read,
\begin{align}
    &\ddot{\delta \psi} + 3 H \dot{\delta \psi} - \left(\frac{R}{3}-\frac{k^2}{a^2}\right)\delta \psi +\frac{1}{3}  \left(2 f_{\phi }-4 V_{\phi }\right) \delta \phi +\frac{1}{3} \psi \delta R+\frac{2}{3} \dot{\delta \phi } \dot{\phi } \label{eqn pert F eom}\\ 
    &~~~~~~~~~~~~~~~~~~~~~~~~~~~~~~~~~~~~~~~~~~~~~
    = \dot{\psi} \left(\dot{\Phi }+3 H\Phi  +3\dot{\Psi }\right) +\Phi  \left(2 \ddot{\psi}+3 H \dot{\psi}+\frac{2}{3} \dot{\phi }^2\right),  \nonumber
    \\
    &\ddot{\delta \phi} +3 H \delta \dot{\phi }+ \left(\frac{k^2}{a^2}-f_{\phi \phi }+2 V_{\phi \phi }\right)\delta \phi -\frac{1}{2} \text{$\delta $R} \psi_{\phi } 
    =\dot{\Phi } \dot{\phi } + \Phi  \left(3 H \dot{\phi }+2 \ddot{\phi }\right) + 3\left( H\Phi  +\dot{\Psi }\right) \dot{\phi },
    \label{eqn pert phi eom}
\end{align}
where $\delta R$ is the perturbation of \eqref{eqn R f}, 
\begin{equation}
    \delta R = \frac{ \delta\psi + \alpha\phi\delta\phi }{A}.
\end{equation}
As usual, we define the standard full comoving curvature perturbation as 
\begin{equation}
    {\cal R} = \Psi - \frac{H}{\rho + P}\delta q,
    \label{eq comoving pert standard}
\end{equation}
where $\delta q$ is the field momentum perturbation. In the longitudinal gauge this can be written in the nice form 
\begin{align}
    {\cal R} = \Psi - \frac{H}{\dot{H}}\left(\dot{\Psi}+ H\Phi\right) = \Psi + \frac{H}{\dot{H}}\left(\frac{ \dot{\delta \psi}+\delta \phi  \dot{\phi }-H\delta \psi -\dot{\psi}\Phi}{2\psi}\right).
    \label{eqn comoving }
\end{align}

 In a closed universe the wave-number will correspond to the eigenfunctions of the Laplace-Beltrami operator on spatial sections \cite{Harrison:1967zza, Martin:2003sf}, given by
\begin{equation}
    \frac{k^2}{a^2} = n(n+2)\frac{K}{a^2},
\end{equation}
where $n>2$ is an integer eigenvalue\footnote{$n=0$ is the homogeneous background and $n=1$ is a gauge choice, so we limit ourselves to $n \geq 2$ \cite{Lehoucq:2002wy}}.

To further analyse the evolution of the perturbations we can combine the perturbed Einstein equations and the equations of motion \eqref{eqn EFE pert energy} - \eqref{eqn pert phi eom} into two second order differential equations. This is achieved by using the relation \eqref{eqn pert non-diag relation} and \eqref{eqn delta rho}-\eqref{eqn delta q} to remove $\Phi, ~\dot{\Phi}, ~ \delta\dot{\phi}$ and $\delta \phi$ respectively. Assuming that $f_{RR}\neq0$, this results in two coupled second order equations, 
\begin{align}
\begin{aligned}
    \delta \ddot{\psi} &+ \left[ 5H +\frac{\dot{\psi}}{\psi}+ \frac{( 2V_{\phi }-  f_{\phi })}{\dot{\phi} }\right]\delta \dot{\psi} 
    - \left[2( 2\dot{H}+ H^2) +\frac{(3\psi H - \dot{\psi})(2V_{\phi }-f_{\phi })}{\psi\dot{\phi}} -\frac{10 H \dot{\psi}}{\psi}  -\frac{2  \ddot{\psi}}{\psi} +\frac{\dot{\psi}^2}{\psi^2} +\frac{1}{3 } \frac{k^2}{a^2} \right] \delta \psi
    \\
    & = \psi\left[4 H -6 \frac{(2V_{\phi }-f_{\phi })}{\dot{\phi}} +6\frac{\dot{\psi}}{\psi}\right]\dot{\Psi}
     +
     2\psi\left[H^2+ \left(\frac{\dot{\psi}}{6\psi}-H\right)\frac{(2V_{\phi } - f_{\phi}) }{ \dot{\phi} } +\frac{2}{3} \frac{(k^2-3 K)}{a^2} 
     + 5 H\frac{\dot{\psi}}{\psi} 
     +\frac{\ddot{\psi}}{\psi}  \right]\Psi
    \end{aligned}
    \label{eqn F_ddot}
\end{align}

\begin{align}
    \begin{aligned}
    \ddot{\Psi} &+ {5H}\dot{\Psi} + \left[ 2 (2\dot{H}+H^2) +\frac{1}{3} \frac{k^2-2 K}{a^2}\right]\Psi= \frac{H}{\psi} \delta \dot{\psi} 
    +\frac{2(  2 H'+H^2)- H \frac{\dot{\psi}}{\psi} +\frac{1}{3} \frac{(2k^2 -3 K)}{a^2}}{ \psi} \delta \psi 
    \end{aligned}
    \label{eqn pert PSI}
\end{align}

\begin{figure}[h!]
    \centering
    \vspace{-0.25cm}
        \includegraphics[width=\textwidth]{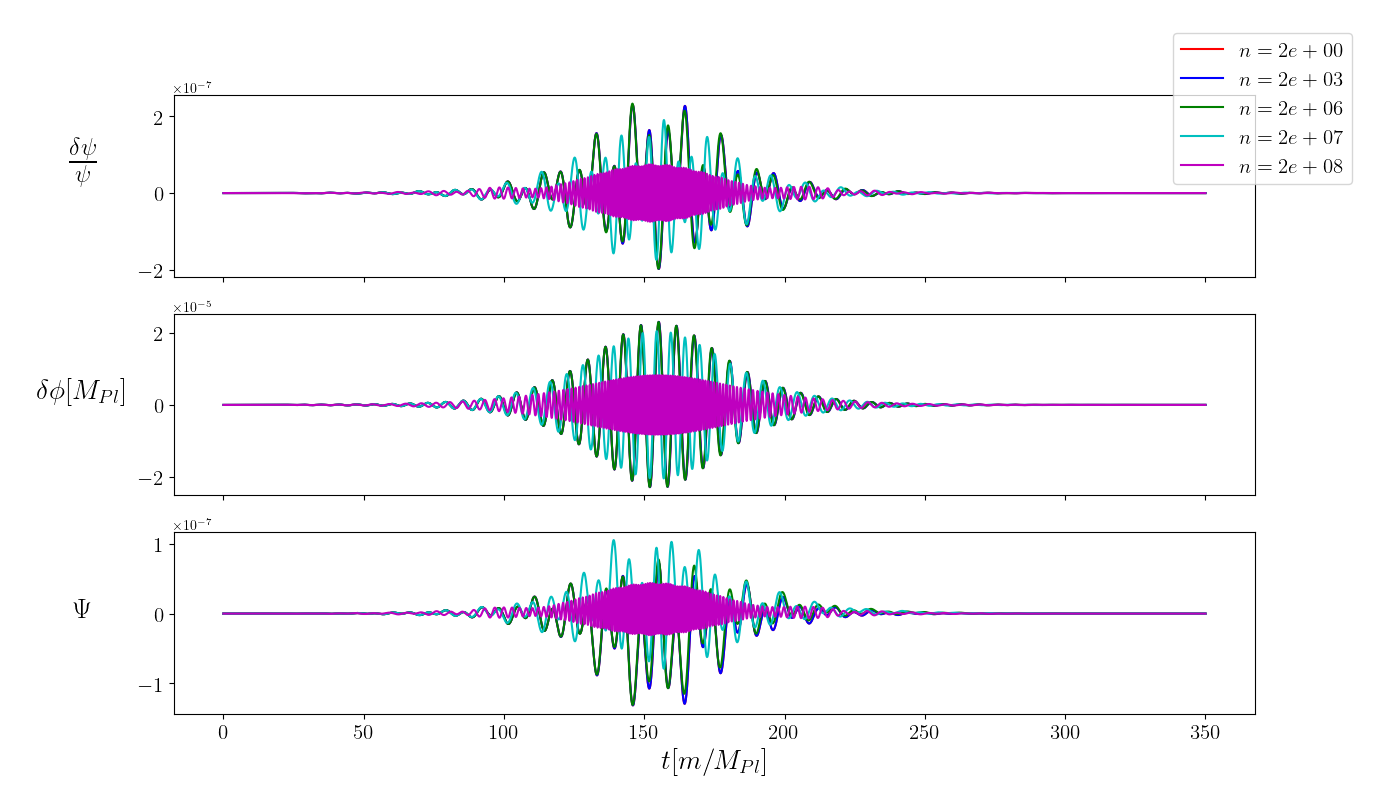}
    \caption{A plot demonstrating the effects of varying $k$. We see that the perturbations stabilise at and tend towards zero as the fields settle back to their initial values. There seems to be no substantial difference in varying the value of $n$. These plots used the same parameter and initial conditions for the background fields as \cref{fig backgound dynamics jordan}.}
    \label{fig pert n eternal}
\end{figure}

\begin{figure}[h!]
    \centering
    \vspace{-0.25cm}
        \includegraphics[width=\textwidth]{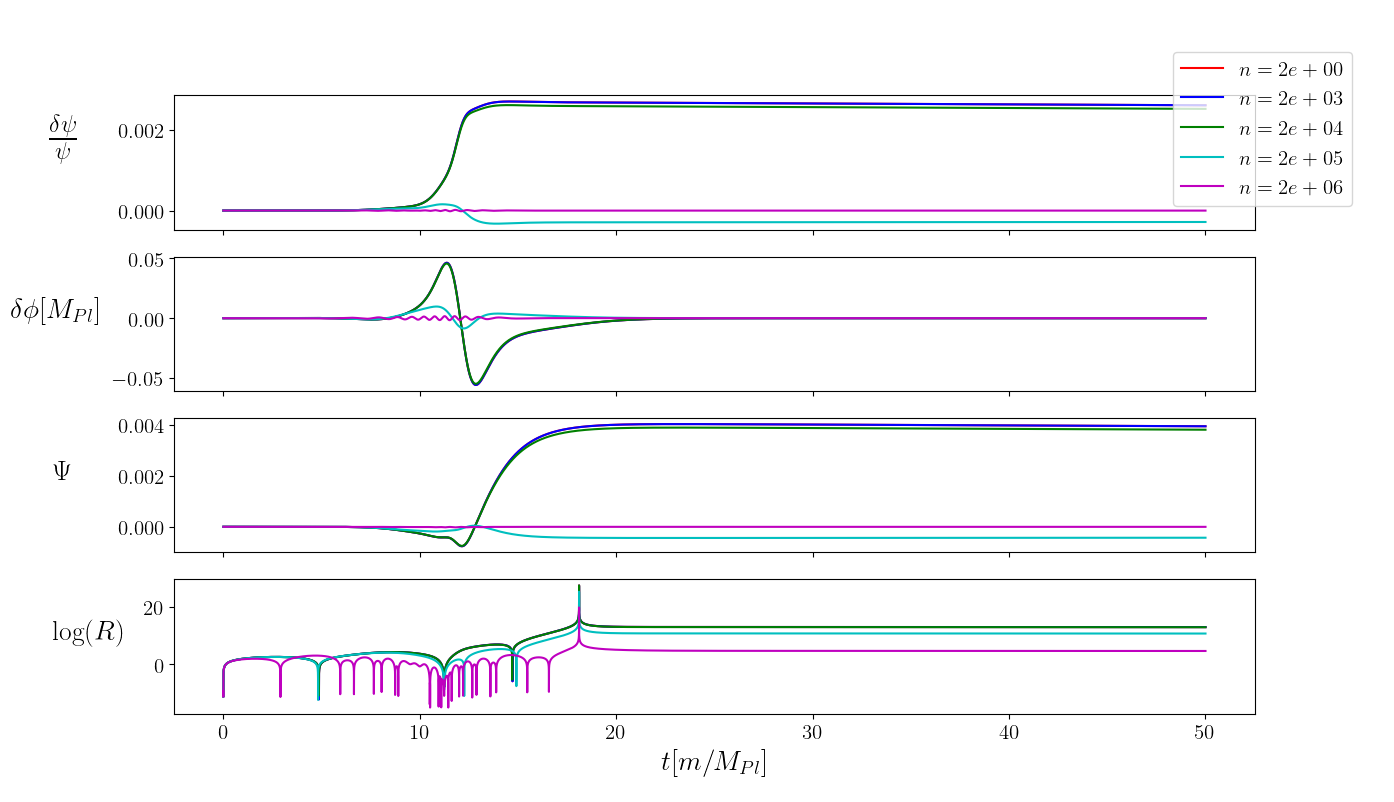}
    \caption{A plot demonstrating the effects of varying $k$. We see that as $\delta\phi\rightarrow0$, $\delta\psi$ stabilises at some constant non-zero value. For $k^2\gg K$ we see that the system is more bounded, with rapid oscillations through the bounce but ultimately tending towards a zero value. These plots used the same parameter and initial conditions for the background fields as \cref{fig background psi_inflaton}. }
    \label{fig pert n}
\end{figure}

However, the scaleron can aid in the stability of $\Psi$. Grouping common terms in \eqref{eqn pert PSI}, we see that within each term there is a counteracting effect between $\delta\psi$ and $\Psi$:  , 
\begin{equation}
    \ddot{\Psi} = - \left( \frac{\delta \dot{\psi}}{\psi} - \frac{\dot{\psi}}{\psi}\frac{\delta \psi}{\psi} - 5\dot{\Psi} \right)|H| - 2\left(2\dot{H}+H^2\right)\left(\Psi - \frac{\delta \psi}{\psi}\right) - \frac{K}{3a^2}\left[\left(n(n+2)-2\right)\Psi - \left(2n(n+2)-3\right)\frac{\delta \psi}{\psi}\right].
\end{equation}
From eqn. \eqref{eqn pert non-diag relation} we expect the two perturbations $\delta\psi/\psi$ and $\Psi$ to have similar magnitudes which is also supported by our numerical results shown in \cref{fig pert n eternal,fig pert n}. In the last equation, terms containing $\delta\psi/\psi$ and $\Psi$ have opposite signs, which implies an counteracting effect, reducing possible divergent behaviour during a collapse in this specific $R^2$ theory.

As discussed further below, for small values of $n$, our numerical calculations show that last two terms will remain negative during the collapse and bouncing epoch. However, the bracket in the first term will remain overall positive during the collapse and bounce, acting as a source term which can lead to an instability. This instability is not present in our simulations, due to the fact that $\delta\dot\psi$ 
increases slower than $\delta\psi$. 

\begin{figure}[h!]
    \centering
    \vspace{-0.25cm}
    \includegraphics[width=\textwidth]{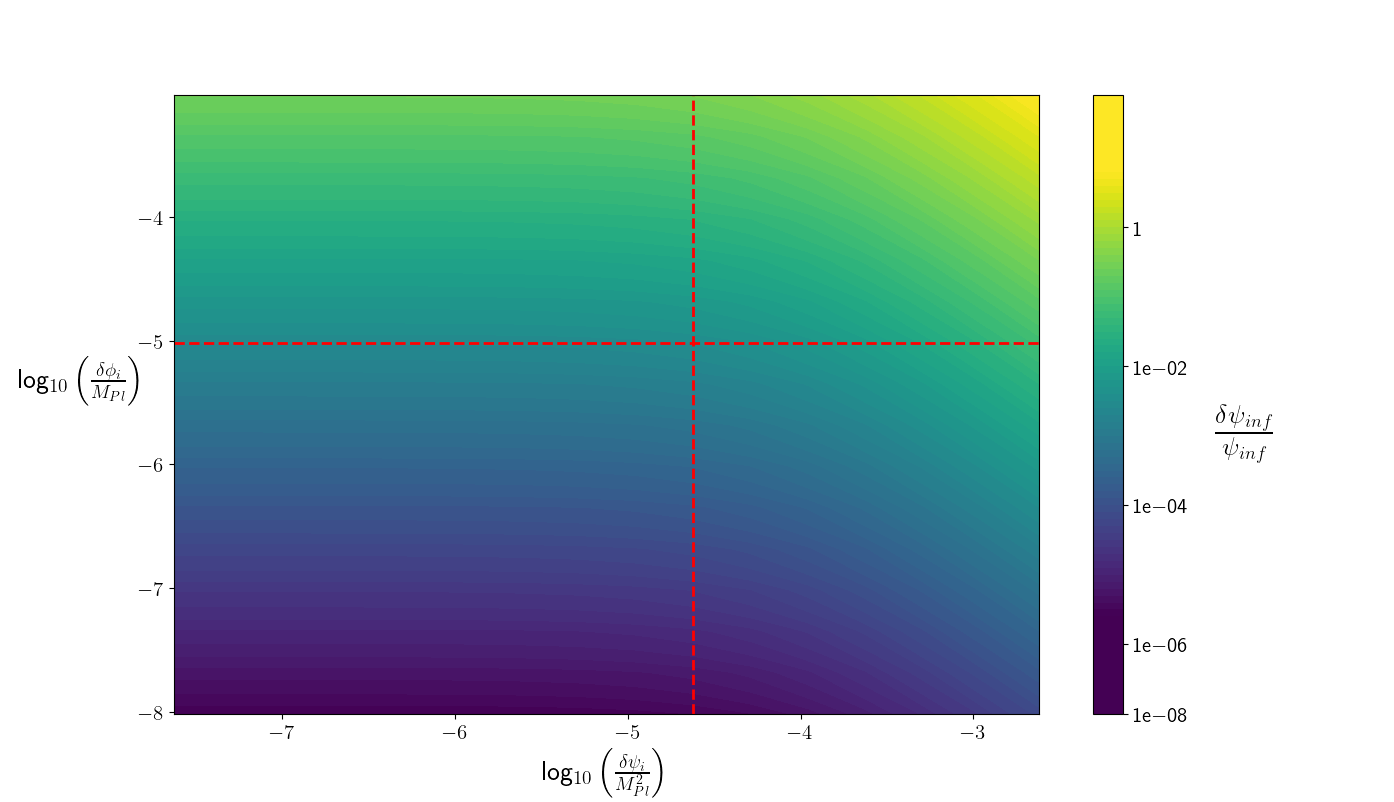}
    \caption{A contour plot illustrating the effect of initial conditions on the perturbations. It is clear there is a large range of initial conditions that lead to stable dynamics. The red line indicates the de-Sitter initial value, $\delta\phi_i = (H/2\pi)$ corresponding to a given $\delta\psi_i$ via eq. \eqref{eqn psi min}.}
    \label{fig pert ic}
\end{figure}
\subsection{Numerical analysis}
The equations just presented are difficult to solve analytically and we therefore resort to a numerical analysis. We numerically integrate \cref{eqn EFE pert momentum,eqn delta q,eqn pert F eom,eqn pert phi eom}, together with \eqref{eqn pert non-diag relation} to remove $\Phi$ and $\dot{\Phi}$.  This also allows us to calculate $\cal R$ from \eqref{eqn comoving }. We use the following initial conditions, unless stated otherwise: $\delta\phi_i = H/(2\pi)$, $\delta\dot{\phi_i} = \dot{H}/(2\pi)$, $\delta \psi_i= \alpha(8AV(\phi_i) -1)\phi_i\delta\phi_i, ~~\delta \dot{\psi_i} = \alpha(8AV(\phi_i) -1)\phi_i\delta\dot{\phi_i}$ from \eqref{eqn psi min},  and setting $\dot{\Psi}=0$. The metric perturbation $\Psi$ is then given by \eqref{eqn EFE pert momentum}\footnote{It we were to set the time derivatives to zero, this would result in a slower growth of the perturbations, resulting in a more stable solution.}. We explore a range of wave-numbers by varying $n$ by orders of magnitude shown in \cref{fig pert n eternal} and \cref{fig pert n}. It is clear that the perturbations increase in amplitude towards the bounce, as expected, but always remain finite. Furthermore the perturbations behave as expected once the inflation regime begins, they settle to a constant value while stretched to superhorizon scales. We analyse the following two scenarios: 

\begin{itemize}
\item\underline{{$\phi\approx\phi_i$} at the bounce:}
This is the case resulting in eternal inflation after the bounce with the resulting numerical perturbations shown in \cref{fig pert n eternal}. As expected the perturbations remain very well behaved, with a slight growth similar to the background field values during the bounce, but ultimately stabilising to zero. A noticeable feature is that a much larger wave number is required to have a noticeable effect, unlike the case for finite inflation shown in \cref{fig pert n}. This is due to the $\phi$-field dominating the dynamics, and has been set to have a lower mass than in the finite inflation case. 

\item\underline{{$0<\phi\lesssim\phi_-$} at the bounce:}
For the scenario which results in finite inflation after the bounce, expressed in \ref{fig pert n}. The wavelength has a prominent effect on the perturbations. Larger wavelengths exhibit a larger growth than perturbations on smaller wavelengths, which is then carried through to the inflationary epoch. 
\end{itemize}

The initial conditions determine the evolution of the perturbations and accordingly whether instabilities are present. We therefore study the stability of the model by changing the initial conditions of the perturbations. Due to the fact $\phi$-field in the finite inflation scenario will always tend towards zero, we only need to analyse the behaviour of the scaleron. This has been numerically verified and is clear from \cref{fig pert n}. If an instability is present, it will manifest a growth in $\delta\psi$. The results are shown in \cref{fig pert ic}. As expected there is a stronger effect of varying the initial condition of $\delta\psi$. However, an interesting feature is that $\delta\phi$ has a noticeable effect on $\delta\psi$. This is a result we have already seen in the background: the scaleron tracking the $\phi$-field. Forcing a larger growth in $\delta\phi$ we also force a larger growth in $\delta\psi$.  

\subsubsection{Vector and Tensor Perturbations}
We briefly discuss the behaviour of vector and tensor perturbations. Let us focus on the vector perturbations first. The vector part of the metric perturbation reads \cite{EllisMaartens,Bari:2019yvk}

\begin{equation}
    ds^2_{(v)} =  -dt^2 + 2a S_i dt dx^i + \left( a^2 \gamma_{ij} + 2 a^2 \partial_j F_i\right) dx^i dx^j~.
\end{equation}
We can define the gauge independent vector shear $\sigma_i = \dot{F_i}+ S_i/a $. The field equations for the vector degrees of freedom read  
\begin{align}
     \delta \dot q_i + 3H\delta q_i &= \frac{k^2 - 2K}{a^2}\delta\Pi_i
    \\
    \frac{k^2 - 2K}{2a^2}\sigma_i & = \frac{\delta q_i}{\psi}~,
\end{align}
where, $\delta \Pi_{ij} = \partial_{(i}\Pi_{j)}$ is the vector part of the anisotropic stress and $\delta q_i$ is the momentum density perturbation. Interestingly only one equation is modified by the additional degree of freedom \cite{Hwang:1996bc}. Assuming that the anisotropic stress vanishes, we can solve the first equation above to give $$ \delta q_i = \delta q_i^{(\rm ini)} \left( \frac{a_{\rm ini}}{a} \right)^3,$$ where the subscript 'ini' denotes the initial values for the scale factor and $\delta q_i$. The second equation above leads then to $$\left(k^2 - 2K \right)\sigma_i = \delta q_i^{(\rm ini)}\frac{a_{\rm ini}^3}{a }\frac{2}{\psi}.$$
As it can be seen from these solutions, the vector perturbations remain small during the collapsing phase as long as the initial amplitude $\delta q_i^{(\rm ini)}$ is small. The scalaron $\psi$ shows an almost exponential behaviour, while $a$ behaves closer to a quadratic centred around the bounce, thus we expect the quantity  $(a\psi)^{-1}$ to peak around the bounce. Numerically we see a growth of two orders of magnitude. After the bounce, the expansion of the universe will cause the vector perturbations to become subdominant, decaying with the expansion of the universe.

\begin{figure}
    \centering
    \includegraphics[width=\textwidth]{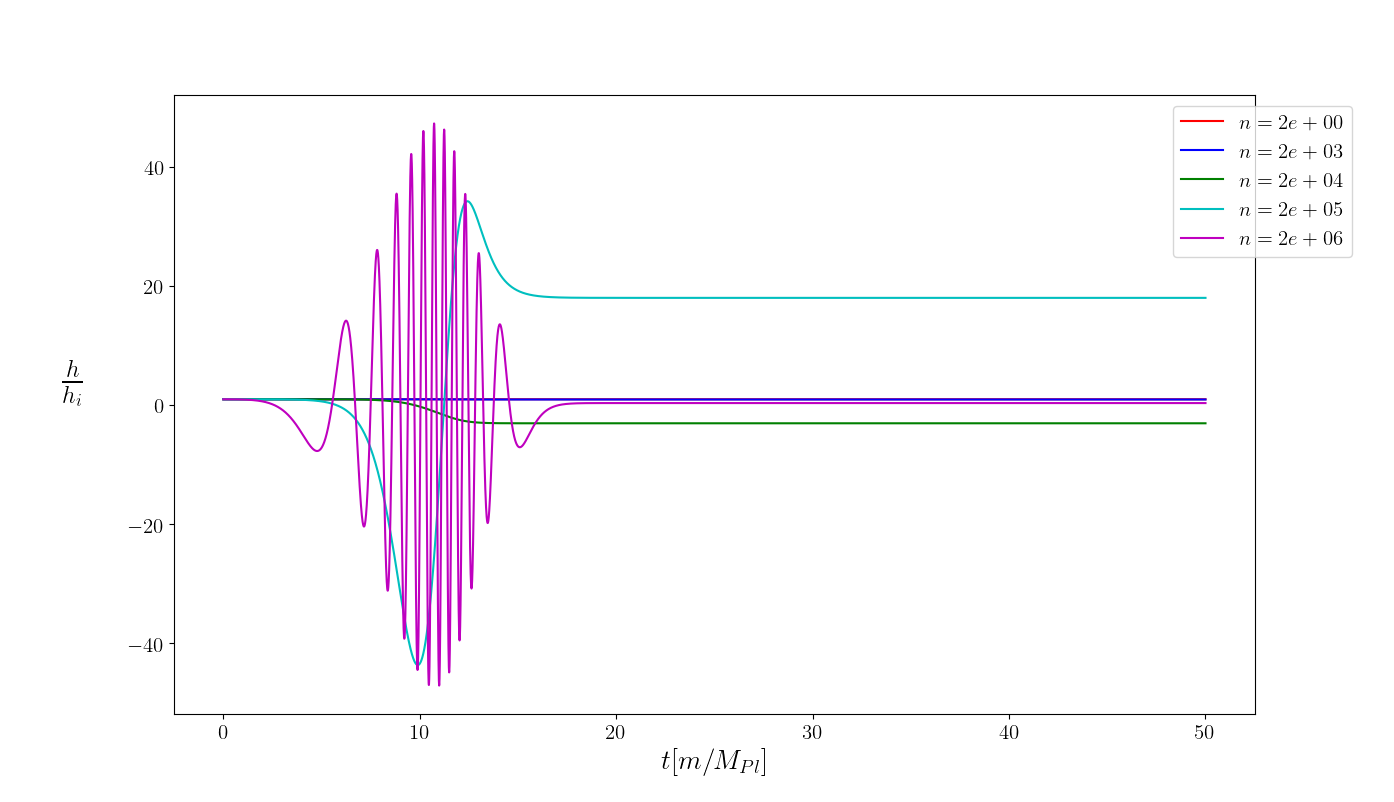}
    \caption{Here we see the amplitude growth of tensor perturbations governed by \cref{eqn pert h} for a range of wavelengths. The parameters used are the same for \cref{fig pert ic}, with initial perturbation chosen to be $h_i = 10^{-5}$.}
    \label{fig pert h}
\end{figure}

Finally we examine the evolution of tensor perturbations $h_{ij}$, defined by 
\begin{equation}
    ds^2_{(T)} = -dt^2 + a^2\left( \gamma_{ij} + h_{ij} \right) dx^i dx^j~.
\end{equation}
In the context of modified gravity and in a closed universe, the gravitational wave equation reads 
\begin{equation}
    \ddot{h} + \left(3H + \frac{\dot{\psi}}{\psi}\right)\dot{h} + \left(\frac{k^2 + 2K}{a^2}\right)h =0, 
    \label{eqn pert h}
\end{equation}
where $h$ is the amplitude of the two polarisation sates, $h_{ij} = h e^{(+,\times)}_{ij}$. 
Provided the scale factor does not vanish, we see that the amplitude will remain finite as illustrated in \cref{fig pert h}. Moreover, the addition of modified gravity reduces the growth of $h$ during the collapse, as $\dot{\psi}/\psi$ counteracts the Hubble term. We therefore conclude that the tensor modes remain small. 

\section{Conclusions}
We studied a classical bouncing universe with a transition into an inflationary phase, using a generalised Starobinsky $f(R)$ model. In order to achieve the bounce without violating the null energy condition, our model assumes a spatially closed universe. Our work builds upon the work conducted by \cite{Gungor:2020fce} to address the initial conditions leading to a bounce: we considered a collapsing universe, dominated by dark energy. The field responsible for dark energy sits in a false vacuum up until about the time the bounce happens. It then settles at the true minimum of the effective potential. Our approach to this model purposefully allows the method to be further generalised for other $f(R,\phi)$ models with false vacuum potentials. In our work the scalaron is responsible for driving a period of inflation after the bounce, which occurs naturally in our setup. The scaleron obtains a kick from $\phi$ during the bounce, driving it up its effective potential, but remaining almost dormant before and during the bounce. Thus, the interaction between the dark energy field and the scalaron leads naturally to an inflationary epoch after the bounce. Moreover, the inclusion of an additional degree of freedom does not lead to an avoidance of the bounce and we find evidence that the scalaron assist stability of the model, as discussed in \cref{sec perturbations}. 

In future work, further details regarding the perturbations will be explored. This includes constraining the model, e.g. calculating the power spectra of perturbations generated before and during inflation. It would be interesting to investigate the features in the primordial power spectra arising the non-standard dynamics before inflation. It will also be interesting to extend the model to study a cyclic cosmology. We address these questions in future work. 

\section*{Acknowledgements:}
We are grateful to Özenç Güngör for interesting discussions.  RD is supported by a STFC CDT studentship. CvdB is supported (in part) by the Lancaster–Manchester–Sheffield Consortium for Fundamental Physics under STFC grant: ST/T001038/1. PD is grateful for support from the First Rand Bank (South Africa). MC is supported by the University of Cape Town Science Faculty Equity Scholarship and the National Research Foundation (NRF South Africa) Scarce-Skills Scholarship.

\vfill

\appendix

\section{Bounce conditions}
\label{sec bounce cond}
In this appendix we briefly discuss the conditions for the bounce and the implications for our model. The necessary conditions for a bounce are (see: \cite{Molina-Paris:1998xmn})
\begin{align*}
    H(t_{b})&=0, & \dot{H}(t_{b})&>0, & \rho+3p&<0. 
\end{align*}
If we assume a barotropic fluid of the form $p=w\rho$, the violation of the strong energy condition constrains the equation of state to $w_{b}<-1/3$. An obvious quantity we can check at the bounce by imposing the first two bounce conditions above is (using Eq. \eqref{eqn Hdot jordan})
\begin{equation}
    \psi_{b}+\frac{\alpha}{2}\phi_{b}^{2} > \frac{6AK}{a_{b}^{2}} + \frac{M_{\rm Pl}^{2}}{2}.
\end{equation}
This equation is fulfilled using the values in \cref{fig backgound dynamics jordan}.
The "slow-roll" parameter $\epsilon = - \dot H / H^2$ is related to the equation of state via 
\begin{equation}
    \epsilon=\frac{3}{2}\left(1+w\right).
\end{equation}
The condition on $\epsilon$ at the bounce is $\epsilon_{b}<1$.\\
Using Eqs. \eqref{eqn H2 jordan}--\eqref{eqn Hdot jordan} and the bounce conditions above, we find the expression for $\epsilon_{b}$:  
\begin{equation}
    \begin{split}
        \epsilon_{b} &= \frac{3}{2}\left(\frac{\rho_{b}+p_{b}}{\rho_{b}}\right)\\
        &= \frac{3A(4K\psi_{b} - a_{b}^{2}\ddot{\psi}_{b} - a_{b}^{2}R_{b}\psi_{b})}{2AK\psi_{b} + \frac{1}{6}a_{b}^{2}\left[\frac{1}{2}(M_{\rm Pl}^{2}-\alpha\phi_{b}^{2}) - \psi_{b}\right]^{2}} < 1
    \end{split}
\end{equation}

\bibliography{references.bib}

\end{document}